\documentclass[preprint,12pt]{elsarticle}

\usepackage{graphicx} 
\usepackage{multirow}%
\usepackage{amsmath,amssymb,amsfonts}%
\usepackage{amsthm}%
\usepackage{mathrsfs}%
\usepackage[title]{appendix}%
\usepackage{xcolor}%
\usepackage{textcomp}%
\usepackage{manyfoot}%
\usepackage{booktabs}%
\usepackage{algorithm}%
\usepackage{algorithmicx}%
\usepackage{algpseudocode}%
\usepackage{listings}%
\usepackage{appendix}

\newcommand{\yb}{$^{171}$Yb$^+$ }
\newcommand{\one}{$|1\rangle$ }

\begin{document}

\begin{frontmatter}

\title{Design and characterization of individual addressing optics based on multi-channel acousto-optic modulator for \yb qubits}

\author[1,2]{Sungjoo Lim}
\author[1]{Seunghyun Baek}
\author[3,4]{Jacob Whitlow}
\author[3,4]{Marissa D\textquotesingle Onofrio}
\author[3,5]{Tianyi Chen}
\author[3,4]{Samuel Phiri}
\author[4,5]{Stephen Crain\fnref{fn1}}
\author[3,4,5]{Kenneth R. Brown}
\author[3,4,5,6]{Jungsang Kim\corref{cor1}}
\ead{jungsang@duke.edu}
\author[1,7]{Junki Kim\corref{cor1}}
\ead{junki.kim.q@skku.edu}

\affiliation[1]{organization={SKKU Advanced Institute of Nanotechnology (SAINT) \& Department of Nano Science and Technology, Sungkyunkwan University},
                city={Suwon},
                postcode={16419}, 
                country={Korea}}
\affiliation[2]{organization={Department of Physics and Astronomy, Seoul National University},
                city={Seoul},
                postcode={08826},
                country={Korea}}
\affiliation[3]{organization={Duke Quantum Center, Duke University},
                city={Durham, NC},
                postcode={27701},
                country={USA}}
\affiliation[4]{organization={Department of Electrical and Computer Engineering, Duke University},
                city={Durham, NC},
                postcode={27708},
                country={USA}}
\affiliation[5]{organization={Department of Physics, Duke University},
                city={Durham, NC},
                postcode={27708},
                country={USA}}                
\affiliation[6]{organization={IonQ, Inc.}, 
                city={College Park, MD},
                postcode={20740},
                country={USA}}
\affiliation[7]{organization={Department of Nano Engineering, Sungkyunkwan University}, 
                city={Suwon}, 
                postcode={16419}, 
                Country={Korea}}
\fntext[fn1]{Present address: IonQ, Inc., College Park, MD 20740}

\cortext[cor1]{Corresponding authors}

\begin{abstract}
    We present the design and characterization of individual addressing optics based on a multi-channel acousto-optic modulator (AOM) for trapped ytterbium-171 ions.
    The design parameters of the individual addressing system were determined based on the tradeoff between the expected crosstalk and the required numerical aperture of the projection objective lens, and we found the optimal target beam diameter and separation to 1.90 \textmu m and 4.28 \textmu m, respectively.
    The individual beams shaped by the projection optics were characterized by two complementary methods: capturing the beams at an intermediate image plane and scanning the ion positions while monitoring the ion-field interactions.
    The resulting effective beam diameters and separations were approximately 2.34--2.36 \textmu m and 4.31 \textmu m, respectively, owing to residual aberration.
    
\end{abstract}

\begin{keyword}
    Ion trap \sep quantum computing \sep quantum information \sep individual addressing optics
\end{keyword}

\end{frontmatter}
\section{Introduction}
Trapped ions are promising quantum information science and technology platforms that exhibit long coherence times \cite{wang_single-qubit_2017, wang_single_2021} and high-precision quantum-state control \cite{ballance_high-fidelity_2016,gaebler_high-fidelity_2016,clark_high-fidelity_2021,wang_high-fidelity_2020, srinivas_high-fidelity_2021}.
Quantum information is encoded in the internal states of trapped ions, and their quantum states can be precisely manipulated using a coherent laser or microwave field.
To independently control each ion's internal state, exposing only the selected ions to the control field is essential and it can be realized by multiple tightly focused control laser beams.
Individual addressing beam arrays for ion qubits have been achieved using multi-channel acousto-optic modulators (AOMs)\cite{debnath_demonstration_2016,wright_benchmarking_2019}, acousto-optic deflectors \cite{pogorelov_compact_2021}, micro-electromechanical system (MEMS) tilting mirrors \cite{crain_individual_2014,wang_high-fidelity_2020}, and waveguide optics \cite{pogorelov_compact_2021, mehta_integrated_2016,mehta_integrated_2020,niffenegger_integrated_2020,binai-motlagh_guided_2023}.
Among these, the multi-channel AOM approach enables independent control of the intensity, frequency, and phase of individual beams and demonstrates parallel entangling gates \cite{figgatt_parallel_2019} and simultaneous sideband cooling \cite{chen_efficient-sideband-cooling_2020}.

In this paper, we report the design and characterization of a multi-channel AOM-based individual addressing optical setup for up to 32 ions.
We studied the effect of design parameters on the intensity crosstalk between neighboring ions and designed projection optics according to the optimized parameters.
The resulting individual beams were monitored using an image sensor in the intermediate imaging plane and characterized using a trapped ion as a field probe.

\section{Setup design}

We define two hyperfine clock states of \yb on a qubit basis, where $|0\rangle$ is $^2$S$_{1/2}$ $|F=0, m_F =0\rangle$ and $|1\rangle$ is $^2$S$_{1/2}$ $|F=1, m_F =0\rangle$.
Coherent transitions between qubit states can be realized through stimulated Raman transitions driven by a mode-locked 355-nm Nd:YAG laser (Coherent Paladin).
To achieve the individual control of ion qubits in a linear chain, switchable individual beams that are tightly focused on each ion are necessary.

We built modular optical systems comprising upstream and downstream optical submodules interconnected by single-mode photonic crystal fibers \cite{colombe_single-mode_2014, spivey_high-stability_2021}.
The upstream module is responsible for distributing the laser beams to downstream modules, modulating the frequencies of laser beams, and switching the laser beams using acoustic-opto modulators (AOMs).
We built two downstream modules to shape and project Raman laser beams onto the \yb ion qubits.
One downstream module was used for global illumination, whereas the other was used for individual addressing beams of up to 32 ions \cite{debnath_programmable_2016, wright_benchmarking_2019}.
Raman transitions of the ions can be realized by either two counter-propagating Raman beams from the downstream modules (counter-propagating configuration) or by a single two-tone Raman beam from one of the downstream modules (co-propagating configuration).
In this study, we focused on the design and characterization of the individual addressing downstream optical submodules.

We employed a 32-channel AOM (L3 Harris) to control the individual Raman beams for up to 32 ions.
The input laser beam of the individual downstream module was divided into 32 separate beams using a diffractive optical element (DOE) and each beam was directed into the 32-channel AOM. 
By controlling the radio frequency (RF) signal applied to each AOM channel, the individual beams diffracted by the AOM can be switched and modulated.
The nominal active aperture size and separation between AOM channels were 140 \textmu m and 450 \textmu m, respectively. 
To avoid UV-induced damage to the optics, the entire AOM and DOE systems were purged with pure nitrogen gas. 

\begin{figure}
    \centering
    \includegraphics[width=\textwidth]{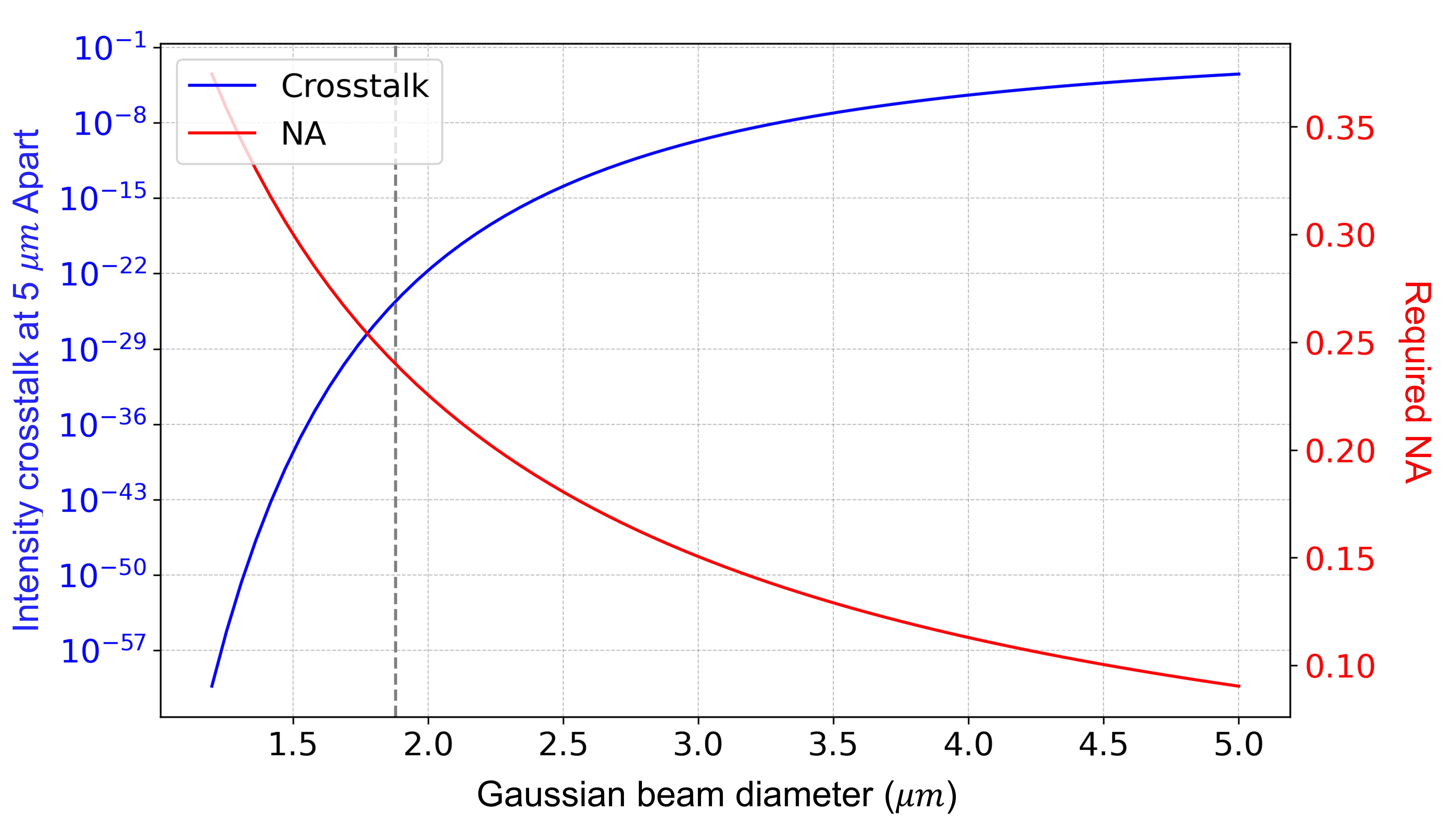}
    \caption{
    Expected intensity crosstalk and required NA as a function of the axial beam size. The vertical dashed line indicates the lower bound of the Gaussian beam diameter determined by the objective lens used in our system (NA=0.24).}
    \label{fig:fig1}
\end{figure}

The following projection optics relayed AOM output beams onto an ion chain with target magnifications.
We used different magnifications along the ion chain axis and perpendicular radial direction because of differing design criteria.
Along the ion chain axis, the size of each beam affects the laser intensity spillover onto neighboring ions and the required numerical aperture (NA) of the focusing objective lens.
Figure \ref{fig:fig1} shows the tradeoff between the expected intensity crosstalk and the required NA of the projection lens.
The intensity crosstalk is defined by the laser intensity spillover normalized by the peak intensity at the nearest ion position, assumed to be 5 \textmu m. 
It induces a gate error in neighboring ions and is minimized at small beam sizes.
However, a small beam requires a high-NA objective lens for tight focusing, making alignment difficult and increasing susceptibility to aberrations.
To minimize the distortion by beam clipping, the NA of the objective lens should be higher than twice of the target Gaussian beam divergence, \textit{i.e.}, NA$> 2\sin(\lambda/\pi w_0)$ where $\lambda$ is the wavelength and $w_0$ is the target beam radius.
With the objective NA of 0.24, we found the minimum diameter of the Gaussian beam to be 1.88 \textmu m and the corresponding minimum crosstalk was $10^{-25}$.

In the radial direction, perpendicular to the ion chain axis, the optical clearance of the employed surface trap was considered.
The optimal beam size along the direction perpendicular to the chip surface can be defined by minimizing the beam clipping on the trap surface.
We used surface traps fabricated by Sandia National Laboratory \cite {maunz_high_2016,revelle_phoenix_2020} and given the geometry of the traps, the optimal Gaussian beam diameter was approximately 5--10 \textmu m.

\begin{figure}
    \centering
    \includegraphics[width=\textwidth]{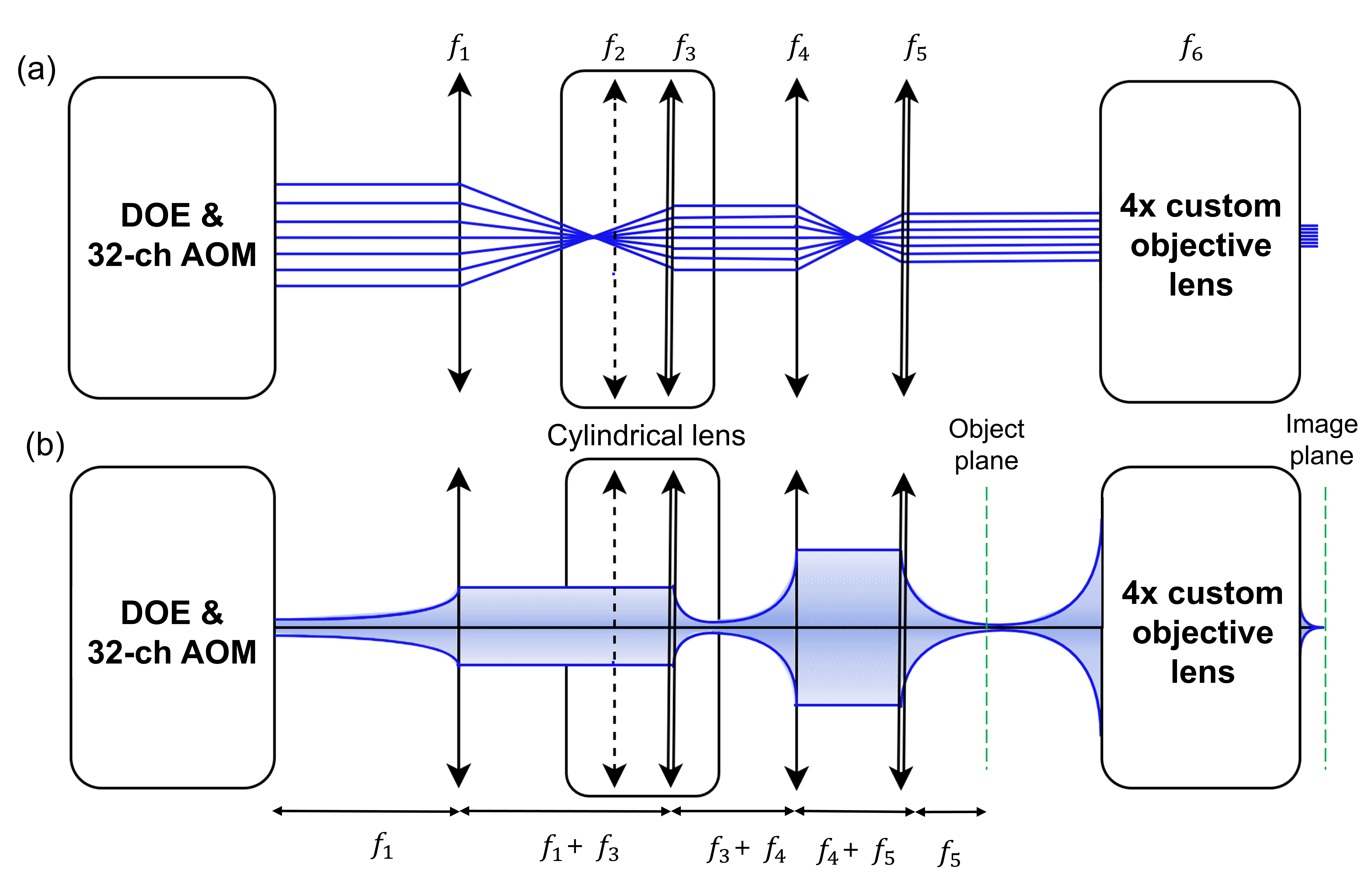}
    \caption{Schematics of the individual beam projection optics. The three-stage telescope relays the beam array launched from the multi-channel AOM to the target ions. The dimensions are not to scale. (a) Trajectory of the individual beam arrays and (b) the Gaussian beam propagation along the ion chain axis. A single-line (double-line) arrow represents a singlet (doublet) lens, and the dashed arrow represents a cylindrical lens along the radial direction of the ion chain. The telescopes are designed for a target magnification of $105\times$ and $21\times$ in the axial and radial directions of the ion chain, respectively.}
    \label{fig:fig2}
\end{figure}

Based on the design criteria above, we built a three-stage telescope system to relay and shape the individual beams (Fig.\ref{fig:fig2}).
The telescope system maps the beams on AOM centers (object plane) to the ion chains (imaging plane).
The first telescope comprised cylindrical lenses with effective focal lengths of 75 mm and 15 mm (a pair of 30-mm lenses) to realize elliptical beams with an aspect ratio of 1:5.
The second telescope has $\sim$2x magnification using spherical lenses, and the final stage used a custom-made 4x magnification objective lens with an imaging NA of 0.24 to minimize spherical aberration.
The beam diameters at the AOM center were adjusted to 200 \textmu m, to match the target beam diameter and beam separations. 
The overall magnifications were $105\times$ along the chain axis and $21\times$ in the radial direction, and the resulting beam diameter at the ion position was expected to be 1.90 \textmu m and 9.58 \textmu m in the axial and radial directions, respectively.
The estimated beam separation was 4.28 \textmu m;
at this distance, there is a sufficient Coulomb interaction between ions for a qubit entangling gate while maintaining an acceptable tradeoff in crosstalk reduction. 
Using a ray-tracing simulation (Zemax Optical Studio), we confirmed that the designed optical system was diffraction-limited.

As the output beams and ion chains are both one-dimensional arrays, it is important to match the beam orientation to the ion chain axis.
Although the beam array from the 32-channel AOM was oriented horizontally with respect to the optical table, the ions were oriented vertically.
To rotate the orientation by $90^{\circ}$, we employed a periscope built with two mirrors whose normal axes are perpendicular to each other.
Fine adjustments of the orientation were made using a dove prism mounted on the optical path.
Before the beam entered the vacuum chamber, a pair of waveplates was installed to control the polarization of the laser beams.

\section{Results \& Discussion}

\begin{figure}
    \centering
    \includegraphics[width=\textwidth]{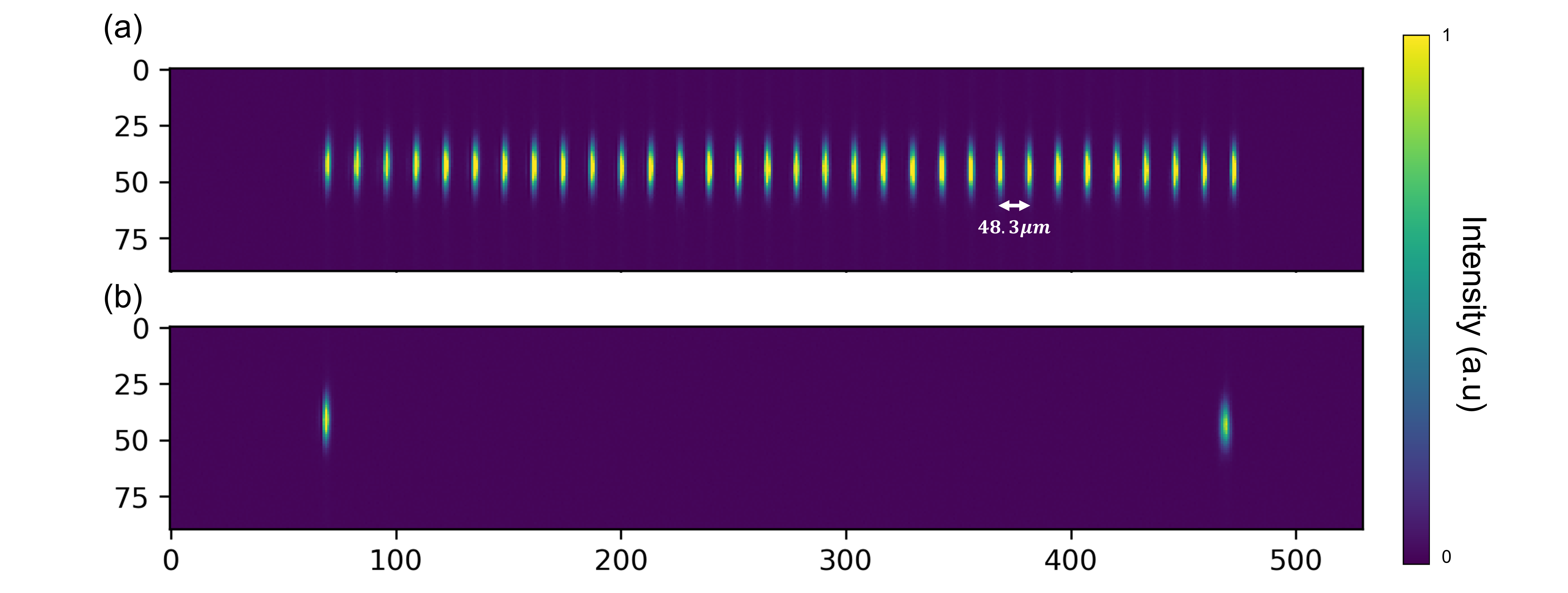}
    \caption{CCD image of the individual beam arrays at an intermediate imaging plane with (a) all 32 channels switched on and (b) only the 1st and 32nd channels on. The average beam separation in the intermediate plane was 48.3 $\pm 0.12$ \textmu m, whereas the expected separation was 44.5 \textmu m.}
    \label{fig:fig3}
\end{figure}

Figure \ref{fig:fig3} shows an intermediate test of the individual beam intensity distributions in the imaging plane of the first telescope.
A CCD image sensor captured the beams after they were directed by a temporary mirror and a unit-magnification telescope.
We confirmed that all 32 beams could be individually switched on or off by toggling RF signals.
The average separation of beams at this intermediate plane was $48.3 \pm 0.12$ \textmu m, whereas that from the simulation result was 44.5 \textmu m.
We believe the slight mismatch of the beam separation and image aberration mostly comes from the temporary unit-magnification telescope.
The relative orientation between the principal axes of the elliptical beams and the beam array axis was matched within the image resolution limit of 15 mrad.

For a more comprehensive analysis, we measured the intensity profiles of individual beams using a trapped \yb ion as a probe. 
When a trapped ion experiences Raman transition in a co-propagation configuration, the Rabi frequency of the transitions is proportional to the laser intensity at the ion position.
By adjusting the DC voltages providing an electrical trapping potential to the ions, we scanned the ion position along the axial direction.
The ion position was monitored using an imaging system at $20\times$ magnification and measurement resolution of 0.19 \textmu m. 

\begin{figure}
    \centering
    \includegraphics[width=\textwidth]{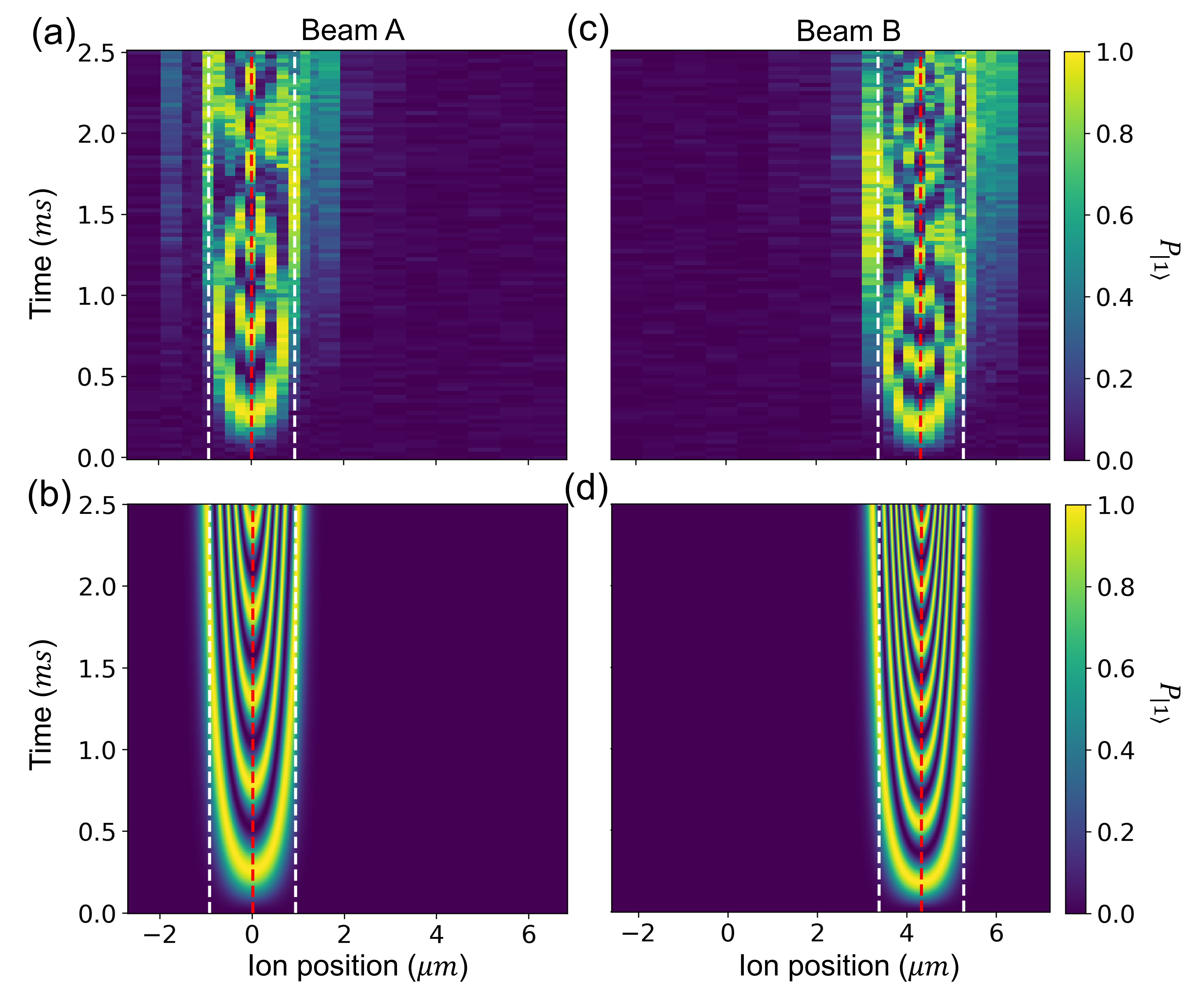}
    \caption{Stimulated Raman transition for various pulse durations and ion positions of two individual beams (a, c) and their model fit (b, d).
    The beam A (B) corresponds to the 16th (17th) beam out of 32 individual beams.
    The red dashed lines show the fit center and the white dashed lines show the Gaussian fit radius. The Gaussian beam diameters of beams A and B were 1.86 \textmu m and 1.88 \textmu m, respectively. }
    \label{fig:fig4}
\end{figure}

Figure \ref{fig:fig4} shows the Rabi oscillations from two nearby Raman beams, and their model fits for various ion positions.
We investigated two central individual beams of the 16th (beam A) and 17th (beam B) out of 32 beams due to the limitations of the established control system.
We applied individual Raman beam pulses with various pulse durations while scanning the ion position along the axial direction and measured the probability of finding the qubit in the \one state.
The expected probability follows a resonant Rabi oscillation trace, and assuming a Gaussian intensity profile for the individual beams, the experimental data were fitted using the following model:
\begin{equation}
    P_{|1\rangle}(x, t) = \sin^2 \left( \frac{\Omega_0 t}{2} \exp(-2(x-x_c)^2/w_0^2 )\right),
\end{equation}
\noindent where $\Omega_0$ is the peak Rabi frequency and $x_c$ and $w_0$ are the Gaussian beam center and diameter, respectively.
We found that the model of fit agreed well with the data, although rapid oscillations were not captured because of the relatively large sampling interval in the ion positions.
From the fitting, the Gaussian diameters of beams A and B were 1.86 \textmu m and 1.88 \textmu m, respectively.
The beam distance was 4.31 \textmu m, which agrees with the simulation within the ion position resolution.
The peak Rabi frequencies of beams A and B were 2$\pi \times$ 1.91 kHz and 2$\pi \times$ 2.79 kHz, respectively, where the intensity imbalance originated from technical issues in the RF sources.

\begin{figure}
    \centering
    \includegraphics[width=\textwidth]{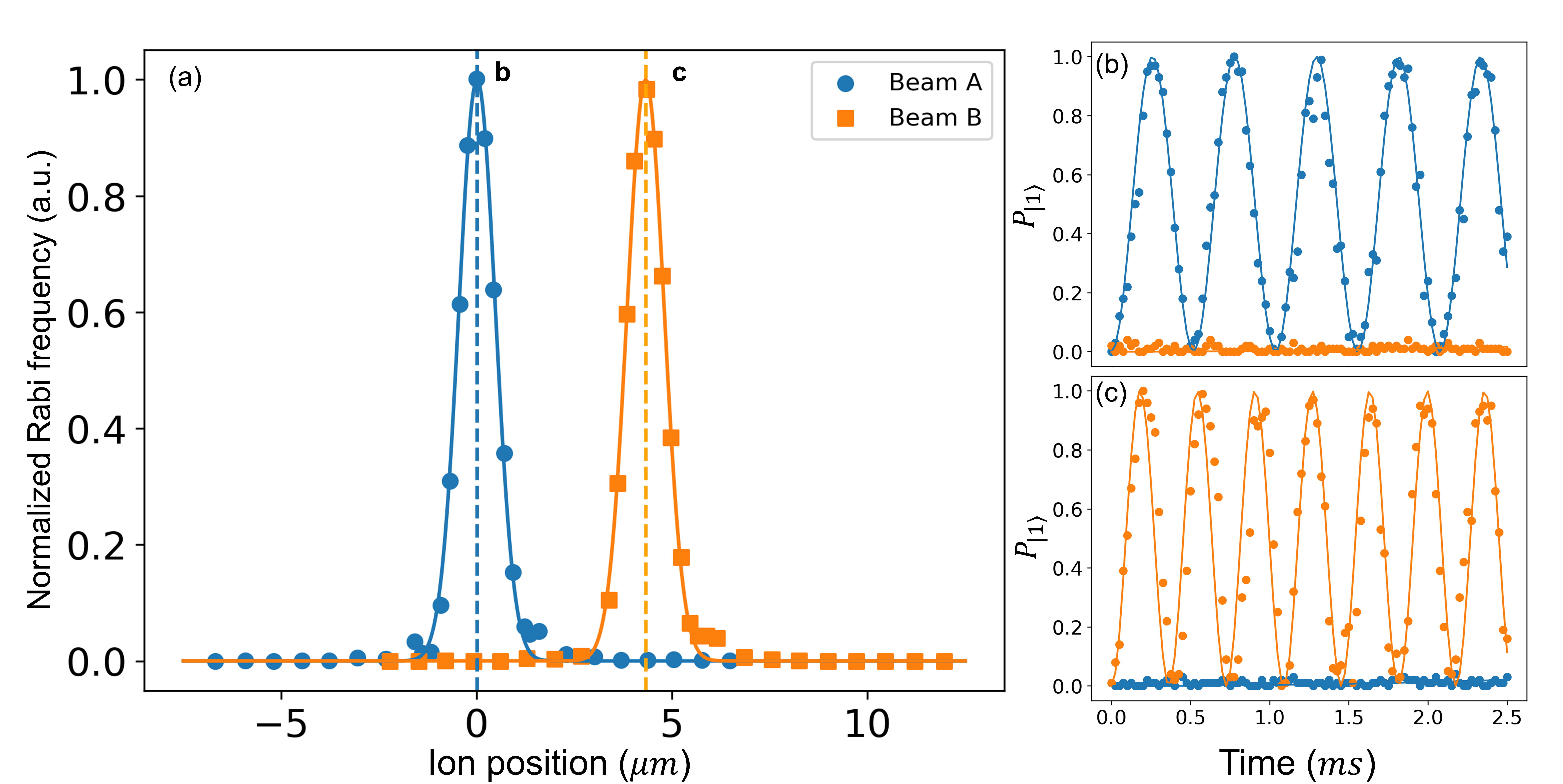}
    \caption{(a) Normalized Rabi frequency vs. ion positions for the two individual beams A and B. Rabi frequencies are normalized by each peak value. The solid line is the normalized Gaussian beam intensity distribution plotted based on the parameters obtained in Fig.\ref{fig:fig4}. (b, c) Rabi oscillation traces at the center of beam A (b) and beam B (c). The solid lines are sine-squared fits of the data. No increase of $P_{|1\rangle}$ was observed within the measured SPAM error, suggesting the intensity crosstalk is less than 0.6\%.}
    \label{fig:fig5}
\end{figure}

We also analyzed the intensity spillover of each Raman beam (Fig. \ref{fig:fig5}).
Figure \ref{fig:fig5} (a) presents the spatial distribution of Rabi frequencies of each beam, where the Rabi frequencies were extracted from the Rabi oscillation traces at each position.
From the achieved spatial Rabi frequency distribution, we calculated the second-moment width (D4$\sigma$) of each beam as 2.34 \textmu m (beam A) and 2.36 \textmu m (beam B), which are approximately 10\% larger than the Gaussiabn team diameter.
This mismatch is believed to be caused by imperfect optical alignment and residual aberrations.

Finally, we compared the Rabi oscillations of the two beams at their peak centers (Fig. 5 (b,c)).
When the system operates, the ions should be pinned at the centers of individual beams, and the effect of neighboring individual beams should be negligible to minimize gate crosstalk.
Compared to the vivid oscillatory trace at the beam center, we did not observe any discernible increase in $P_{|1\rangle}$ within 2.5 ms due to influence from the neighboring beams.
This absence of influence from neighboring beams suggests the effective suppression of crosstalk from neighboring beams.
With a state preparation and measurement error of 1\%, the data indicate an upper limit of the crosstalk Rabi frequency at 2$\pi \times$ 12.8 Hz, corresponding to only 0.6\% of the intensity crosstalk to the nearest ion qubits.

In conclusion, we designed and constructed an individual addressing optical setup for Raman transitions of \yb qubits. 
The design parameters were determined to suppress the gate errors induced by intensity crosstalk and minimize beam aberration.
The individual beams were characterized using both intermediate CCD images and stimulated Raman transitions while scanning the ion position, which agreed with prediction and suggested an intensity crosstalk of less than 0.6\%.
The tailored individual addressing module demonstrated the capability of providing reliable and programmable operations for universal quantum computation.

\clearpage

\section*{Declaration of competing interest}

The authors declare that they have no known competing financial interests or personal relationships that could have appeared to influence the work reported in this paper.

\section*{Acknowledgments}

This work was supported by the National Research Foundation of Korea (NRF) grant funded by the Korea government(MSIT) (2021M3H3A103657312, 2022M3E4A105136512, 2022M3K4A109712211, 2022M3H3A106307412) and National Science Foundation STAQ Project, PHY-1818914.


\begin{thebibliography}{10}
\expandafter\ifx\csname url\endcsname\relax
  \def\url#1{\texttt{#1}}\fi
\expandafter\ifx\csname urlprefix\endcsname\relax\def\urlprefix{URL }\fi
\expandafter\ifx\csname href\endcsname\relax
  \def\href#1#2{#2} \def\path#1{#1}\fi

\bibitem{wang_single-qubit_2017}
Y.~Wang, M.~Um, J.~Zhang, S.~An, M.~Lyu, J.-N. Zhang, L.-M. Duan, D.~Yum, K.~Kim, \href{https://www.nature.com/articles/s41566-017-0007-1}{Single-qubit quantum memory exceeding ten-minute coherence time}, Nature Photonics 11~(10) (2017) 646--650, number: 10 Publisher: Nature Publishing Group.
\newblock \href {https://doi.org/10.1038/s41566-017-0007-1} {\path{doi:10.1038/s41566-017-0007-1}}.
\newline\urlprefix\url{https://www.nature.com/articles/s41566-017-0007-1}

\bibitem{wang_single_2021}
P.~Wang, C.-Y. Luan, M.~Qiao, M.~Um, J.~Zhang, Y.~Wang, X.~Yuan, M.~Gu, J.~Zhang, K.~Kim, \href{https://www.nature.com/articles/s41467-020-20330-w}{Single ion qubit with estimated coherence time exceeding one hour}, Nature Communications 12~(1) (2021) 233, number: 1 Publisher: Nature Publishing Group.
\newblock \href {https://doi.org/10.1038/s41467-020-20330-w} {\path{doi:10.1038/s41467-020-20330-w}}.
\newline\urlprefix\url{https://www.nature.com/articles/s41467-020-20330-w}

\bibitem{ballance_high-fidelity_2016}
C.~Ballance, T.~Harty, N.~Linke, M.~Sepiol, D.~Lucas, \href{https://link.aps.org/doi/10.1103/PhysRevLett.117.060504}{High-{Fidelity} {Quantum} {Logic} {Gates} {Using} {Trapped}-{Ion} {Hyperfine} {Qubits}}, Physical Review Letters 117~(6) (2016) 060504, publisher: American Physical Society.
\newblock \href {https://doi.org/10.1103/PhysRevLett.117.060504} {\path{doi:10.1103/PhysRevLett.117.060504}}.
\newline\urlprefix\url{https://link.aps.org/doi/10.1103/PhysRevLett.117.060504}

\bibitem{gaebler_high-fidelity_2016}
J.~Gaebler, T.~Tan, Y.~Lin, Y.~Wan, R.~Bowler, A.~Keith, S.~Glancy, K.~Coakley, E.~Knill, D.~Leibfried, D.~Wineland, \href{https://link.aps.org/doi/10.1103/PhysRevLett.117.060505}{High-{Fidelity} {Universal} {Gate} {Set} for \$\{{\textasciicircum}\{9\}{\textbackslash}mathrm\{{Be}\}\}{\textasciicircum}\{+\}\$ {Ion} {Qubits}}, Physical Review Letters 117~(6) (2016) 060505, publisher: American Physical Society.
\newblock \href {https://doi.org/10.1103/PhysRevLett.117.060505} {\path{doi:10.1103/PhysRevLett.117.060505}}.
\newline\urlprefix\url{https://link.aps.org/doi/10.1103/PhysRevLett.117.060505}

\bibitem{clark_high-fidelity_2021}
C.~R. Clark, H.~N. Tinkey, B.~C. Sawyer, A.~M. Meier, K.~A. Burkhardt, C.~M. Seck, C.~M. Shappert, N.~D. Guise, C.~E. Volin, S.~D. Fallek, H.~T. Hayden, W.~G. Rellergert, K.~R. Brown, \href{https://link.aps.org/doi/10.1103/PhysRevLett.127.130505}{High-{Fidelity} {Bell}-{State} {Preparation} with \${\textasciicircum}\{40\}\{{\textbackslash}mathrm\{{Ca}\}\}{\textasciicircum}\{+\}\$ {Optical} {Qubits}}, Physical Review Letters 127~(13) (2021) 130505, publisher: American Physical Society.
\newblock \href {https://doi.org/10.1103/PhysRevLett.127.130505} {\path{doi:10.1103/PhysRevLett.127.130505}}.
\newline\urlprefix\url{https://link.aps.org/doi/10.1103/PhysRevLett.127.130505}

\bibitem{wang_high-fidelity_2020}
Y.~Wang, S.~Crain, C.~Fang, B.~Zhang, S.~Huang, Q.~Liang, P.~H. Leung, K.~R. Brown, J.~Kim, \href{https://link.aps.org/doi/10.1103/PhysRevLett.125.150505}{High-{Fidelity} {Two}-{Qubit} {Gates} {Using} a {Microelectromechanical}-{System}-{Based} {Beam} {Steering} {System} for {Individual} {Qubit} {Addressing}}, Physical Review Letters 125~(15) (2020) 150505, publisher: American Physical Society.
\newblock \href {https://doi.org/10.1103/PhysRevLett.125.150505} {\path{doi:10.1103/PhysRevLett.125.150505}}.
\newline\urlprefix\url{https://link.aps.org/doi/10.1103/PhysRevLett.125.150505}

\bibitem{srinivas_high-fidelity_2021}
R.~Srinivas, S.~C. Burd, H.~M. Knaack, R.~T. Sutherland, A.~Kwiatkowski, S.~Glancy, E.~Knill, D.~J. Wineland, D.~Leibfried, A.~C. Wilson, D.~T.~C. Allcock, D.~H. Slichter, \href{https://www.nature.com/articles/s41586-021-03809-4}{High-fidelity laser-free universal control of trapped ion qubits}, Nature 597~(7875) (2021) 209--213, number: 7875 Publisher: Nature Publishing Group.
\newblock \href {https://doi.org/10.1038/s41586-021-03809-4} {\path{doi:10.1038/s41586-021-03809-4}}.
\newline\urlprefix\url{https://www.nature.com/articles/s41586-021-03809-4}

\bibitem{debnath_demonstration_2016}
S.~Debnath, N.~M. Linke, C.~Figgatt, K.~A. Landsman, K.~Wright, C.~Monroe, \href{https://www.nature.com/articles/nature18648}{Demonstration of a small programmable quantum computer with atomic qubits}, Nature 536~(7614) (2016) 63--66.
\newline\urlprefix\url{https://www.nature.com/articles/nature18648}

\bibitem{wright_benchmarking_2019}
K.~Wright, K.~M. Beck, S.~Debnath, J.~M. Amini, Y.~Nam, N.~Grzesiak, J.-S. Chen, N.~C. Pisenti, M.~Chmielewski, C.~Collins, K.~M. Hudek, J.~Mizrahi, J.~D. Wong-Campos, S.~Allen, J.~Apisdorf, P.~Solomon, M.~Williams, A.~M. Ducore, A.~Blinov, S.~M. Kreikemeier, V.~Chaplin, M.~Keesan, C.~Monroe, J.~Kim, \href{https://www.nature.com/articles/s41467-019-13534-2}{Benchmarking an 11-qubit quantum computer}, Nature Communications 10~(1) (2019) 5464.
\newblock \href {https://doi.org/10.1038/s41467-019-13534-2} {\path{doi:10.1038/s41467-019-13534-2}}.
\newline\urlprefix\url{https://www.nature.com/articles/s41467-019-13534-2}

\bibitem{pogorelov_compact_2021}
I.~Pogorelov, T.~Feldker, C.~D. Marciniak, L.~Postler, G.~Jacob, O.~Krieglsteiner, V.~Podlesnic, M.~Meth, V.~Negnevitsky, M.~Stadler, B.~Höfer, C.~Wächter, K.~Lakhmanskiy, R.~Blatt, P.~Schindler, T.~Monz, \href{https://link.aps.org/doi/10.1103/PRXQuantum.2.020343}{Compact {Ion}-{Trap} {Quantum} {Computing} {Demonstrator}}, PRX Quantum 2~(2) (2021) 020343.
\newblock \href {https://doi.org/10.1103/PRXQuantum.2.020343} {\path{doi:10.1103/PRXQuantum.2.020343}}.
\newline\urlprefix\url{https://link.aps.org/doi/10.1103/PRXQuantum.2.020343}

\bibitem{crain_individual_2014}
S.~Crain, E.~Mount, S.~Baek, J.~Kim, \href{https://doi.org/10.1063/1.4900754}{Individual addressing of trapped {171Yb}+ ion qubits using a microelectromechanical systems-based beam steering system}, Applied Physics Letters 105~(18) (2014) 181115.
\newblock \href {https://doi.org/10.1063/1.4900754} {\path{doi:10.1063/1.4900754}}.
\newline\urlprefix\url{https://doi.org/10.1063/1.4900754}

\bibitem{mehta_integrated_2016}
K.~K. Mehta, C.~D. Bruzewicz, R.~McConnell, R.~J. Ram, J.~M. Sage, J.~Chiaverini, \href{https://www.nature.com/articles/nnano.2016.139}{Integrated optical addressing of an ion qubit}, Nature Nanotechnology 11~(12) (2016) 1066--1070, number: 12 Publisher: Nature Publishing Group.
\newblock \href {https://doi.org/10.1038/nnano.2016.139} {\path{doi:10.1038/nnano.2016.139}}.
\newline\urlprefix\url{https://www.nature.com/articles/nnano.2016.139}

\bibitem{mehta_integrated_2020}
K.~K. Mehta, C.~Zhang, M.~Malinowski, T.-L. Nguyen, M.~Stadler, J.~P. Home, \href{https://www.nature.com/articles/s41586-020-2823-6}{Integrated optical multi-ion quantum logic}, Nature 586~(7830) (2020) 533--537, number: 7830 Publisher: Nature Publishing Group.
\newblock \href {https://doi.org/10.1038/s41586-020-2823-6} {\path{doi:10.1038/s41586-020-2823-6}}.
\newline\urlprefix\url{https://www.nature.com/articles/s41586-020-2823-6}

\bibitem{niffenegger_integrated_2020}
R.~J. Niffenegger, J.~Stuart, C.~Sorace-Agaskar, D.~Kharas, S.~Bramhavar, C.~D. Bruzewicz, W.~Loh, R.~T. Maxson, R.~McConnell, D.~Reens, G.~N. West, J.~M. Sage, J.~Chiaverini, \href{https://www.nature.com/articles/s41586-020-2811-x}{Integrated multi-wavelength control of an ion qubit}, Nature 586~(7830) (2020) 538--542, number: 7830 Publisher: Nature Publishing Group.
\newblock \href {https://doi.org/10.1038/s41586-020-2811-x} {\path{doi:10.1038/s41586-020-2811-x}}.
\newline\urlprefix\url{https://www.nature.com/articles/s41586-020-2811-x}

\bibitem{binai-motlagh_guided_2023}
A.~Binai-Motlagh, M.~Day, N.~Videnov, N.~Greenberg, C.~Senko, R.~Islam, \href{http://arxiv.org/abs/2302.14711}{A guided light system for agile individual addressing of {Ba}\${\textasciicircum}+\$ qubits with \$10{\textasciicircum}\{-4\}\$ level intensity crosstalk}, arXiv:2302.14711 [physics, physics:quant-ph] (Feb. 2023).
\newblock \href {https://doi.org/10.48550/arXiv.2302.14711} {\path{doi:10.48550/arXiv.2302.14711}}.
\newline\urlprefix\url{http://arxiv.org/abs/2302.14711}

\bibitem{figgatt_parallel_2019}
C.~Figgatt, A.~Ostrander, N.~M. Linke, K.~A. Landsman, D.~Zhu, D.~Maslov, C.~Monroe, \href{https://www.nature.com/articles/s41586-019-1427-5}{Parallel entangling operations on a universal ion-trap quantum computer}, Nature 572~(7769) (2019) 368--372.
\newblock \href {https://doi.org/10.1038/s41586-019-1427-5} {\path{doi:10.1038/s41586-019-1427-5}}.
\newline\urlprefix\url{https://www.nature.com/articles/s41586-019-1427-5}

\bibitem{chen_efficient-sideband-cooling_2020}
J.-S. Chen, K.~Wright, N.~C. Pisenti, D.~Murphy, K.~M. Beck, K.~Landsman, J.~M. Amini, Y.~Nam, \href{https://link.aps.org/doi/10.1103/PhysRevA.102.043110}{Efficient-sideband-cooling protocol for long trapped-ion chains}, Physical Review A 102~(4) (2020) 043110, publisher: American Physical Society.
\newblock \href {https://doi.org/10.1103/PhysRevA.102.043110} {\path{doi:10.1103/PhysRevA.102.043110}}.
\newline\urlprefix\url{https://link.aps.org/doi/10.1103/PhysRevA.102.043110}

\bibitem{colombe_single-mode_2014}
Y.~Colombe, D.~H. Slichter, A.~C. Wilson, D.~Leibfried, D.~J. Wineland, \href{https://opg.optica.org/abstract.cfm?uri=oe-22-16-19783}{Single-mode optical fiber for high-power, low-loss {UV} transmission}, Optics express 22~(16) (2014) 19783--19793.
\newline\urlprefix\url{https://opg.optica.org/abstract.cfm?uri=oe-22-16-19783}

\bibitem{spivey_high-stability_2021}
R.~F. Spivey, I.~V. Inlek, Z.~Jia, S.~Crain, K.~Sun, J.~Kim, G.~Vrijsen, C.~Fang, C.~Fitzgerald, S.~Kross, \href{https://ieeexplore.ieee.org/abstract/document/9606562/}{High-stability cryogenic system for quantum computing with compact packaged ion traps}, IEEE Transactions on Quantum Engineering 3 (2021) 1--11.
\newline\urlprefix\url{https://ieeexplore.ieee.org/abstract/document/9606562/}

\bibitem{debnath_programmable_2016}
S.~Debnath, A {Programmable} {Five} {Qubit} {Quantum} {Computer} {Using} {Trapped} {Atomic} {Ions}, Ph.D. thesis, University of Maryland (2016).

\bibitem{maunz_high_2016}
P.~Maunz, \href{https://www.osti.gov/servlets/purl/1237003/}{High {Optical} {Access} {Trap} 2.0.}, Tech. Rep. SAND–2016-0796R, 1237003, 618951 (Jan. 2016).
\newblock \href {https://doi.org/10.2172/1237003} {\path{doi:10.2172/1237003}}.
\newline\urlprefix\url{https://www.osti.gov/servlets/purl/1237003/}

\bibitem{revelle_phoenix_2020}
M.~C. Revelle, \href{http://arxiv.org/abs/2009.02398}{Phoenix and {Peregrine} {Ion} {Traps}} (Sep. 2020).
\newblock \href {https://doi.org/10.48550/arXiv.2009.02398} {\path{doi:10.48550/arXiv.2009.02398}}.
\newline\urlprefix\url{http://arxiv.org/abs/2009.02398}

\end{thebibliography}

\end{document}